\documentclass[%
 aip,
 jmp,%
 amsmath,amssymb,
reprint,%
]{revtex4-1}
\usepackage{graphicx}
\usepackage{dcolumn}
\usepackage{bm}

\begin{document}

\preprint{AIP/123-QED}

\title{Spin dynamics and frequency dependence of magnetic damping study in soft ferromagnetic FeTaC film with a stripe domain structure}

\author{B. Samantaray$^{1a)}$, Akhilesh K. Singh$^{2}$, A.Perumal$^{2}$, R. Ranganathan$^{1}$ and P. Mandal}
\affiliation{Saha Institute of Nuclear Physics, 1/AF Bidhannagar, Calcutta 700 064, India}
\email{iitg.biswanath@gmail.com}
\affiliation{Department of Physics, Indian Institute of Technology Guwahati, Guwahati - 781039, India}

\date{\today}
\begin{abstract}
Perpendicular magnetic anisotropy (PMA) and low magnetic damping are the key factors for the free layer magnetization switching by spin transfer torque technique in magnetic tunnel junction devices. The magnetization precessional dynamics in soft ferromagnetic FeTaC thin film with a stripe domain structure was explored in broad band frequency range by employing micro-strip ferromagnetic resonance technique. The polar angular variation of resonance field  and linewidth at different frequencies have been analyzed numerically using Landau-Lifshitz-Gilbert equation by taking into account the total free energy density of the film. The numerically estimated parameters Land\'{e} $g$-factor, PMA constant, and effective magnetization are found to be 2.1, 2$\times10^{5}$ erg/cm$^{3}$ and 7145 Oe, respectively. The frequency dependence of Gilbert damping parameter ($\alpha$) is evaluated by considering both intrinsic and extrinsic effects into the total linewidth  analysis. The value of $\alpha$ is found to be 0.006 at 10 GHz and it increases with decreasing precessional frequency.

\vskip 1cm
\end{abstract}

\maketitle
\newpage
Spin transfer torque (STT) has grater credibility compared to other techniques towards ultrafast spin dynamics in ferromagnet  by electric current induced magnetization reversal of spin valves and magnetic tunnel junctions (MTJ).\cite{Sankey2008} The current researchers are more keen to focus on STT technology for its high density magnetic random access memories (MRAM),\cite{Devolder2008,Huai2004} STT-driven domain wall devices\cite{Fukami2009} and perpendicular magnetic recording media\cite{Khizroev2004} applications. In order to make this technology more efficient, lowering the critical current density is essential which requires the material specifications with low saturation magnetization ($M_{S}$), high spin polarization,  large uniaxial perpendicular magnetic anisotropy (PMA) constant and low magnetic damping.\cite{Mizukami2013,Song2013,Guo2014} The magnetic damping parameter ($\alpha$) can be described well by the phenomenological Landau-Lifshitz-Gilbert equation and is known as the Gilbert damping.\cite{Landau,Gilbert} Several attempts have been made for understanding the origin of Gilbert damping in spin dynamics relaxation in single layer as well as multilayered magnetic alloys, which arises from both intrinsic and extrinsic parts of the material. The intrinsic contribution to the Gilbert damping parameter has been studied by tuning the strength of the spin-orbit coupling.\cite{He2013,Guo2014,Woltersdorf2009} Recently, Ikeda {\it et al}.\cite{Ikeda2010} have reported that CoFeB-MgO based MTJ with PMA would be reliable for high-density non-volatile memory application due to its high thermal stability and efficiency towards STT technology. The investigation on magnetic dynamics, PMA and the apparent magnetic damping have been studied extensively in CoFeB based soft ferromagnetic thin film by ferromagnetic resonance (FMR) and time-resolved magneto-optical Kerr effect.\cite{Hirayama2015,Lu2014} Malinowski {\it et al}. \cite{Malinowski2009} have reported  a large increase in Gilbert damping with applied magnetic field in perpendicularly magnetized CoFeB thin film.

In this letter, we focus on amorphous FeTaC layer due to its interesting soft ferromagnetic (FM) properties.\cite{Tanahashi2003,Akhilesh2012} The amorphous soft FM layer reduces the number of pinning centers which may lead to the STT-driven domain wall motion along with high tunneling magnetoresistance ratio (TMR). The transcritical loop along with the stripe domain structure, which are the manifestation of PMA component were reported on FeTaC thin film with thickness of 200 nm.\cite{Akhilesh2012,Akhilesh2013} To shed some more light onto its dynamic magnetic properties, we have further studied this film by using ferromagnetic resonance technique. Though the magnetic anisotropy and Gilbert damping have been studied by FMR technique in several magnetic thin films like Heusler alloys, permalloy, soft magnetic materials and multilayered (FM/antiferromagnetic or non-magnetic/FM) magnetic films for magnetic recording, MTJ and TMR reader applications, most of the reports are limited to single frequency due to the measurements in X-band electron-spin-resonance spectrometer where the cavity resonates at particular frequency.\cite{Zakeri2007,Gong2009,Pandey2012,Mizukami2013,Guo2014,Behera2015,Lu2014} In this report, spin dynamics and magnetic relaxation are studied at different magnetization precessional frequencies.

Soft ferromagnetic  Fe$_{80}$Ta$_{8}$C$_{12}$ single layer film with thickness 200 nm was deposited by dc magnetron sputtering technique and the details of growing environment were reported earlier.\cite{Akhilesh2012} The static and dynamic magnetic properties were explored by using a vector network analyzer (VNA) based homemade micro-strip ferromagnetic resonance (MS-FMR) spectrometer. The micro-strip line which was coupled to VNA and Schottky diode detector (Agilent 8473D) through high frequency coaxial cables was mounted in between the pole pieces of the electromagnet. The magnetic thin film with film side downward was mounted on the strip line. The frequency of the microwave signal was fixed by using an Agilent Technologies made VNA (Model PNA-X, N-5242A) with a constant microwave power of 5 dBm. The first derivative of the absorption spectrum with respect to magnetic field ($H$) was collected by field modulation and lock-in detection technique. The FM thin film was treated in-plane and out-of-plane orientations. The magnetic field sweeping FMR spectra were recorded by varying two parameters: precessional frequency and the angle between $H$ and normal of the film. The frequency ($f$) and polar angle ($\theta_{H}$) dependence of resonance field ($H_{r}$) and linewidth ($\Delta H_{PP}$) were extracted from each FMR spectrum and the numerical calculations were carried out by mathematica program for different relaxation processes.

\begin{figure}[t]
\begin{center}
\includegraphics[trim = 0mm 0mm 0mm 10mm, width=50mm]{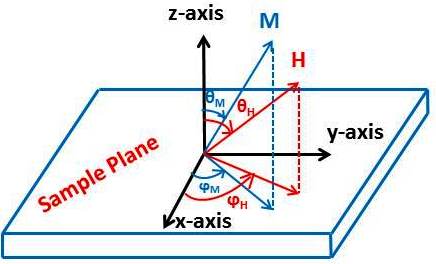}\
\caption{Schematic diagram of $M$ and $H$ vectors in spherical polar coordinate system. $\varphi_{M}$ and $\varphi_{H}$ are the in-plane angle of  magnetization ($M$) and external magnetic field ($H$) with respect to $x$ axis, while $\theta_{M}$ and $\theta_{H}$ are out-of-plane angles with respect to $z$ axis.}\label{F1}
\end{center}
\end{figure}

The precession of magnetization (${M}$) in the sample plane under the influence of microwave and external magnetic field is illustrated in Fig. 1 in a polar coordinate system. $\varphi_{H}$ ($\varphi_{M}$) is the in-plane angle between $H$($M$) and $x$ axis and $\theta_{H}$($\theta_{M}$) is the polar angle between $z$ axis and $H$($M$). The uniform precession of magnetization can be described by the Landau-Lifshitz-Gilbert (LLG) equation of motion,\cite{Landau,Gilbert}

\begin{equation}
\frac{{\partial \overrightarrow M }}{{\partial t}} =  - \gamma \left( {\overrightarrow M  \times {{\overrightarrow H }_{eff}}} \right) + \frac{G}{{\gamma M_S^2}}\left[ {\overrightarrow M  \times \frac{{\partial \overrightarrow M }}{{\partial t}}} \right]\
\end{equation}
The first term corresponds to the precessional torque in the effective magnetic field and the second term is the Gilbert damping torque. $\gamma=g\mu_{B}/\hbar$ is denoted as gyromagnetic ratio and written in terms of Land\'{e} $g$ factor, Bohr magneton $\mu_{B}$, and Planck constant $\hbar$. $G=\gamma\alpha M_{S}$ is related to the intrinsic relaxation rate of the material. $\alpha$ is the dimensionless Gilbert damping parameter. The free energy density of a single magnetic thin film can be written as,

\begin{equation}
\begin{array}{c}
E =  - {M_S}H\left[ {\sin {\theta _H}\sin {\theta _M}cos\left( {{\varphi _H} - {\varphi _M}} \right) + \cos {\theta _H}\cos {\theta _M}} \right]\\
 - 2\pi M_S^2{\sin ^2}{\theta _M} + {K_ \bot }{\sin ^2}{\theta _M}
\end{array}\
\end{equation}
where the first term is analogous to the Zeeman energy, the second term is dipolar demagnetization energy, the third term signifies the anisotropy energy, $M_{S}$ is the saturation magnetization, $K_{\bot}$ is the PMA constant with corresponding anisotropic field $H_{\bot}=2K_{\bot}/M_{S}$. The resonance frequency $f_{r}$ of the uniform precession mode is deduced from the energy density by using the following expression,\cite{Acher2003}

\begin{equation}
f_r^2 = {\left( {\frac{\gamma }{{2\pi }}} \right)^2}\frac{1}{{M_S^2{{\sin }^2}{\theta _M}}}\left[ {\frac{{{\partial ^2}E}}{{\partial \theta _M^2}}\frac{{{\partial ^2}E}}{{\partial \varphi _M^2}} - {{\left( {\frac{{{\partial ^2}E}}{{\partial {\theta _M}\partial {\varphi _M}}}} \right)}^2}} \right]\
\end{equation}
where the derivatives are evaluated at equilibrium positions of $M$ and $H$.

\begin{figure}[t]
\begin{center}
\includegraphics[trim = 0mm 0mm 0mm 0mm, width=86mm]{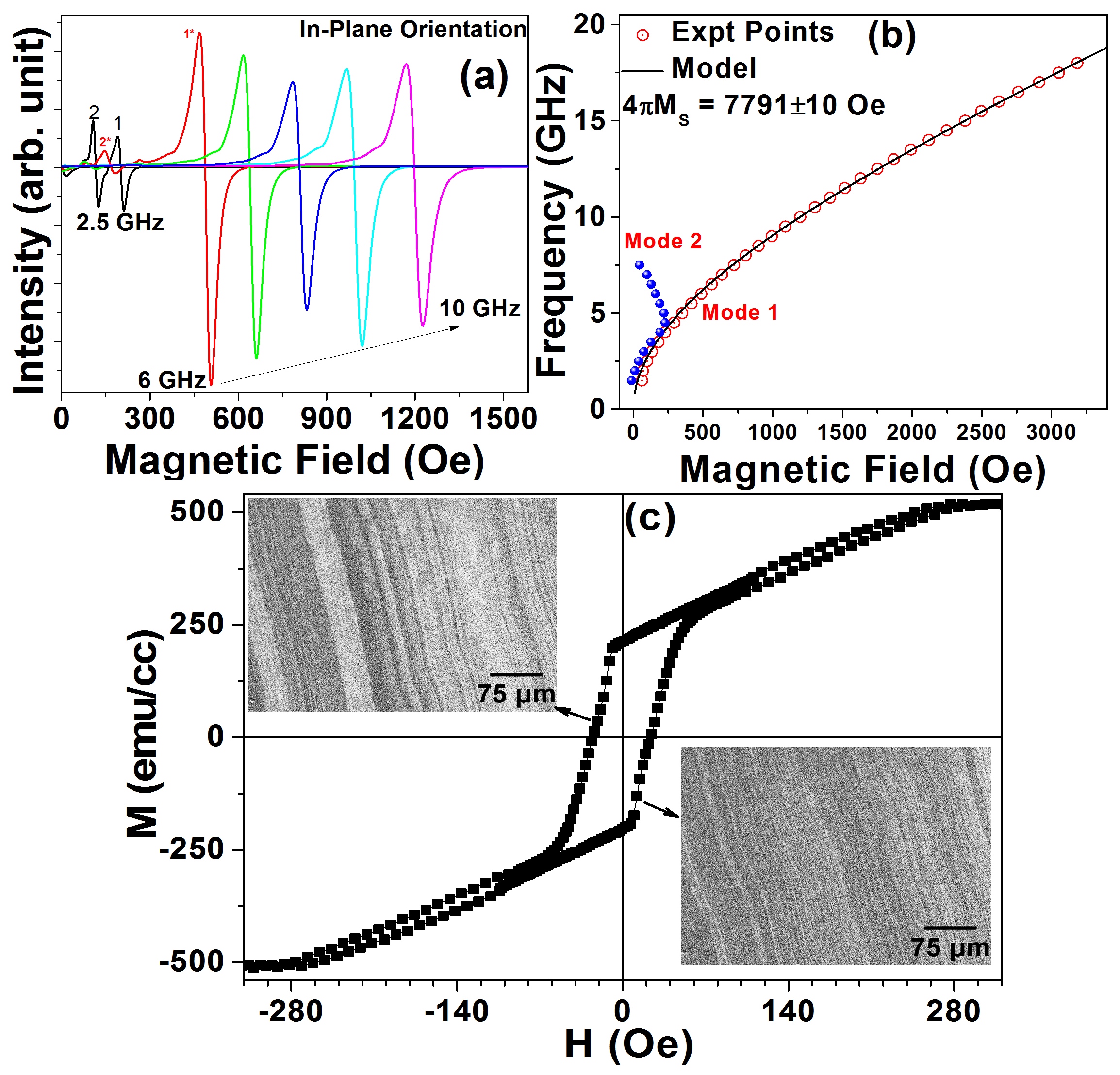}\
\caption{(a) shows typical FMR spectra at different frequencies in planar orientation, (b) shows $f$ dependence of $H_{r}$ for in-plane applied magnetic field. Experimentally and numerically calculated $H_{r}$ values are shown as open circles and solid line respectively and (c) shows the plot of room temperature $M-H$ loop and domain images reproduced from Ref. [19].}\label{F2}
\end{center}
\end{figure}

For the in-plane orientation, the typical FMR spectra  at different frequencies are shown in Fig. 2(a). The measurements were carried out by varying the frequency from 1 to 18 GHz with an interval of 0.5 GHz. In the lower frequency range 1-6 GHz, the FMR spectra show two resonance peaks. The low-field resonance peaks named as secondary mode and are marked by 2 and 2$^{*}$ for 2.5 and 6 GHz, respectively in Fig. 2(a). This mode arises from the linear unsaturated zone of the transcritical $M$($H$) loop \cite{Akhilesh2013} and is usually observed in stripe-domain structure.\cite{Acher1997,Vukadinovic2001} The transcritical loop along with the domain structure are reproduced from earlier report\cite{Akhilesh2013} and is shown in Fig. 2(c). The dense stripe-domain structure observed in this film confirms the presence of perpendicular magnetic anisotropy. The primary modes usually called uniform mode are marked as 1 and 1$^{*}$ for 2.5 and 6 GHz, respectively. The value of $H_{r}$ for secondary resonance peak increases with the increase in $f$ up to 4.5 GHz and then follows the reverse trend as depicted in Fig. 2(b). This could be explained on the basis that the value of $H_{r}$ of the uniform mode above 4.5 GHz overcomes the parallel saturation field, i.e., 280 Oe as observed in $M$($H$) curve. Above 6 GHz, $H_{r}$ exceeds the parallel saturation field in large extent and this could be the reason for the strong attenuation of secondary phase. In planar configuration ($\theta_{M}$=$\theta_{H}$=$\pi/2$), the solution for the in-plane resonance frequency can be calculated by incorporating the total energy in Eq. 3 and is given by,

\begin{equation}
{f_r} = \frac{\gamma }{{2\pi }}{\left[ {\left( {4\pi M + Hcos\left( {{\varphi _H} - {\varphi _M}} \right)} \right)\left( {Hcos\left( {{\varphi _H} - {\varphi _M}} \right)} \right)} \right]^{\frac{1}{2}}}\
\end{equation}
The value of $\varphi_{M}$ can be calculated by using the solution of $H$ at equilibrium condition, i.e., $\frac{{\partial E}}{{\partial {\varphi _M}}}$$=$0. However, for the present thin film, we could not find any planar anisotropy from the $\varphi_{H}$ dependence of $H_{r}$ and hence conclude, $\varphi_{H}$ = $\varphi_{M}$. The $f$ dependence of $H_{r}$ is numerically calculated by using Eq. 4 and is shown as a solid line in Fig. 2(b). The numerically calculated values yielded a good fit and the parameters are found to be reliable with $4\pi M_{S}$ = 7791$\pm$10 Oe and $\gamma$ = 2.95 MHz/Oe with a $g$-factor of 2.1. The deduced value of saturation magnetization is very close to earlier reported value from $M$($H$) loop measurement.\cite{Akhilesh2012}


In out-of-plane configuration, the solution for the resonance frequency is deduced from Eq. 3 by employing the conditions, $\varphi_{H}$=$\varphi_{M}$=$0^{\circ}$ and is represented in Eq. 5.

\begin{equation}
\begin{array}{c}
f_r^2 = {\left( {\frac{\gamma }{{2\pi }}} \right)^2}\left[ {H\cos \left( {{\theta _M} - {\theta _H}} \right) - \left( {4\pi {M_S} - \frac{{2{K_ \bot }}}{{{M_S}}}} \right)\cos 2{\theta _M}} \right]\\
\left[ {H\cos \left( {{\theta _M} - {\theta _H}} \right) - \left( {4\pi {M_S} - \frac{{2{K_ \bot }}}{{{M_S}}}} \right){{\cos }^2}{\theta _M}} \right]
\end{array}\
\end{equation}

\begin{figure}[t]
\begin{center}
\includegraphics[trim = 5mm 0mm 0mm 0mm, width=80mm]{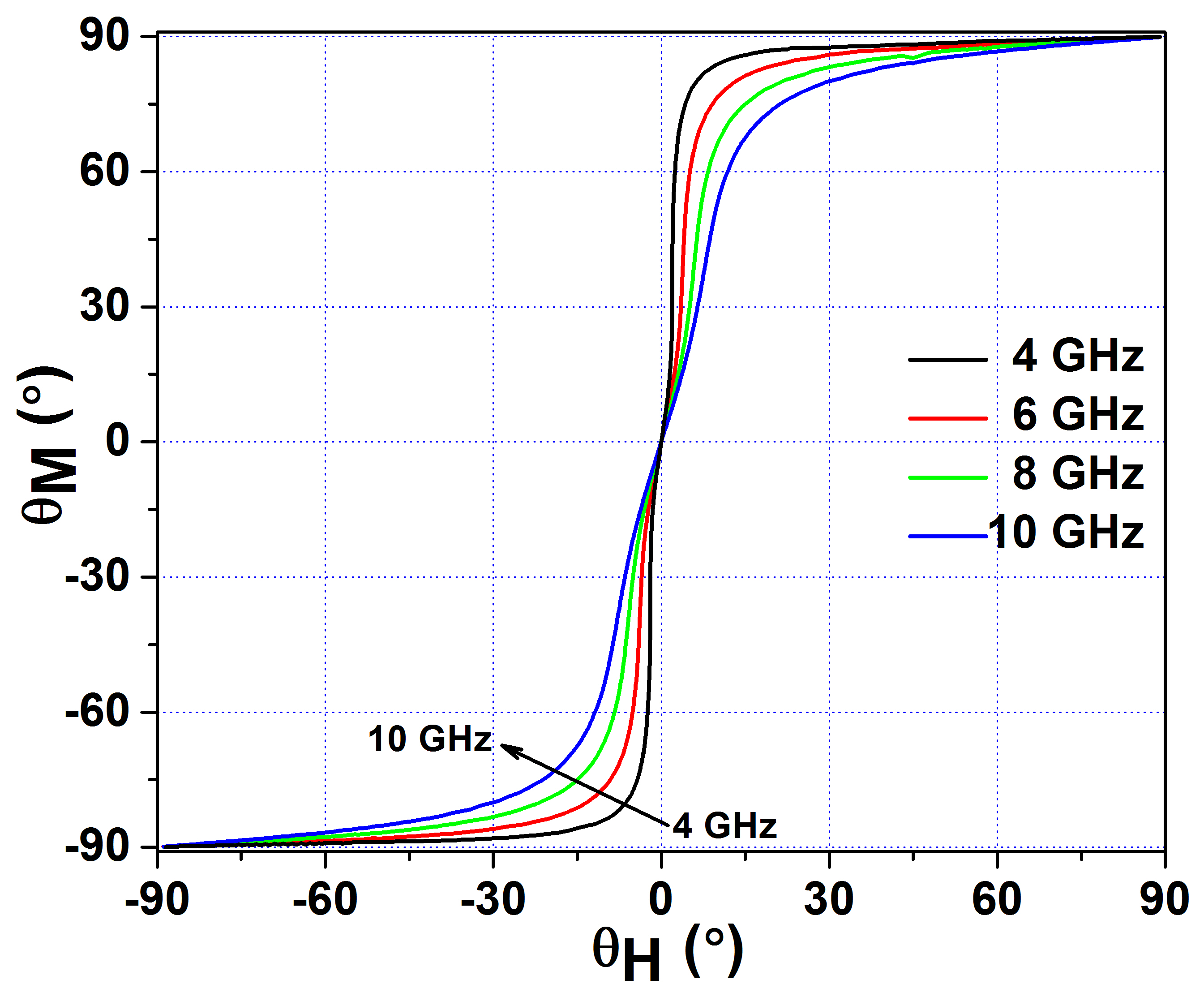}\
\caption{Equilibrium angle of the magnetization, $\theta_{M}$, as a function of the applied field direction, $\theta_{H}$, in out-of-plane configuration at different frequencies.}\label{F3}
\end{center}
\end{figure}

\begin{figure}[h!tb]
\begin{center}
\includegraphics[trim = 5mm 0mm 0mm 0mm, width=80mm]{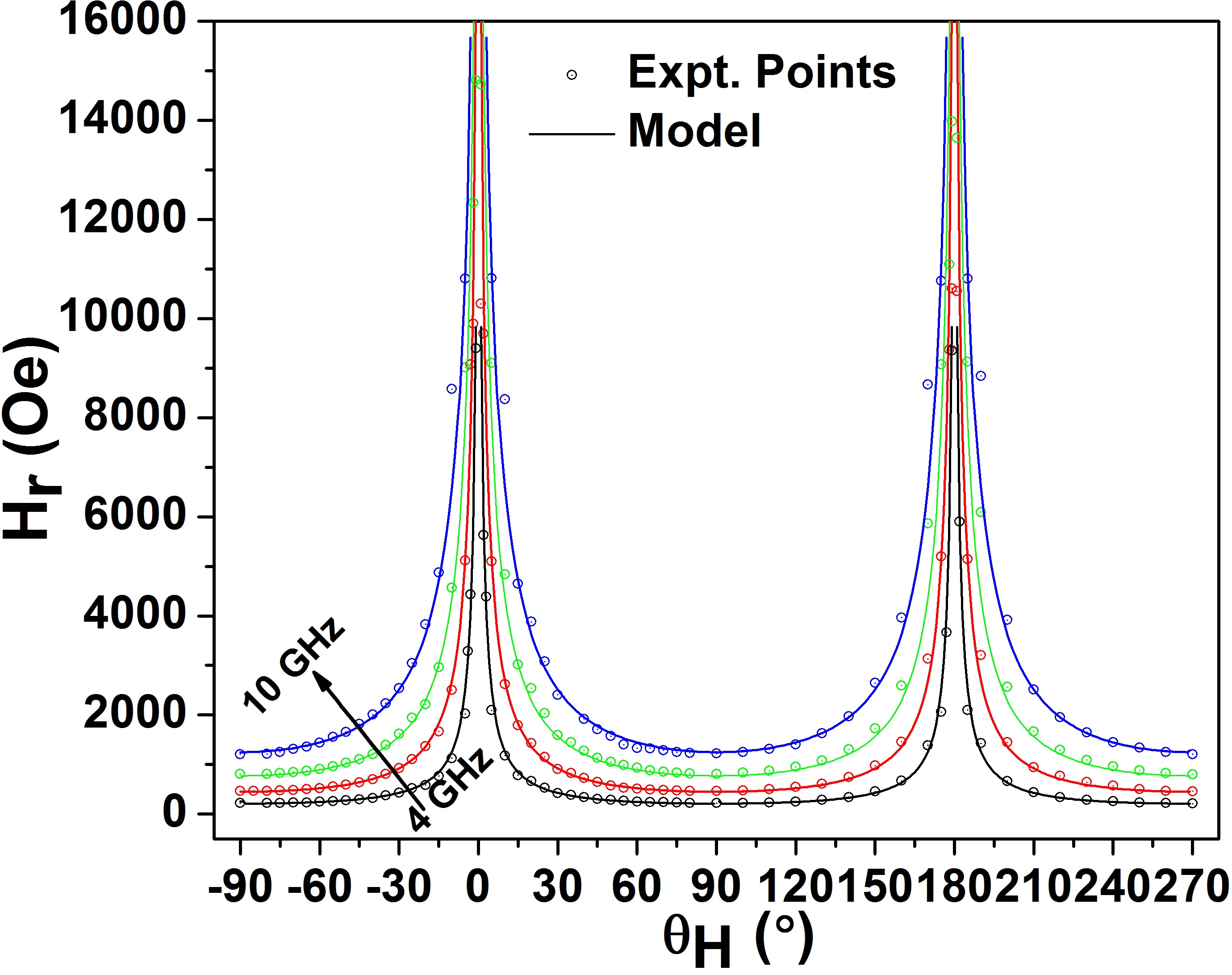}\
\caption{Angular dependence of resonance field $H_{r}$ in out-of-plane configuration at different frequencies. ($\circ$) shows the experimental points and the line (-) shows the modeled data.}\label{F4}
\end{center}
\end{figure}
The equilibrium angle $\theta_{M}$ is numerically calculated for each value of $\theta_{H}$ by minimizing the energy, i.e., $\frac{{\partial E}}{{\partial {\theta_M}}} = 0$ and is depicted in Fig. 3 for different frequencies. Fig. 3 demonstrates that magnetization suddenly attempts to align in planar direction as the magnetic field goes away from the $\theta_{H}$=0$^{\circ}$ and 180$^{\circ}$. Fig. 4 shows one complete round of $\theta_{H}$ dependence of $H_{r}$ at different frequencies. Uniaxial PMA is found to be observed along with singularity at $\theta_{H}=0^{\circ}$ and $\theta_{H}=180^{\circ}$, which signifies that infinite magnetic field is required to turn the $M$ vector parallel to  $H$ in perpendicular configuration.  The dependence of $H_{r}$ on $\theta_{H}$ is modeled at different frequencies starting from 4 to 10 GHz with 2 GHz intervals by using Eq. 5 and the interpolated values of $\theta_{M}$ from Fig. 3. The modeled values of $H_{r}$ are plotted as a solid line in Fig. 4 and a very good agreement with experimental data is observed. The parameters deduced from this calculation are found to be $K_{\bot}$$=$$2\times10^{5}$ erg/cm$^{3}$ and 4$\pi M_{eff}$$=$7145 Oe.


Finally, the damping of magnetization precession has been analyzed from linewidth of FMR spectra. The $\theta_{H}$ dependence of $\Delta H_{PP}$ as shown in Fig. 5 was extracted from the polar angle variation of FMR spectrum at different frequencies in the range of 4-10 GHz with an interval of 2 GHz. In order to get better clarity of Fig. 5, the data for the 4 GHz frequency are not shown. The total linewidth broadening due to the intrinsic and extrinsic parts of the material has been expressed in the following equation,\cite{Mizukami2002,Guo2014}

\begin{figure}[t]
\begin{center}
\includegraphics[trim = 0mm 0mm 0mm 0mm, width=80mm]{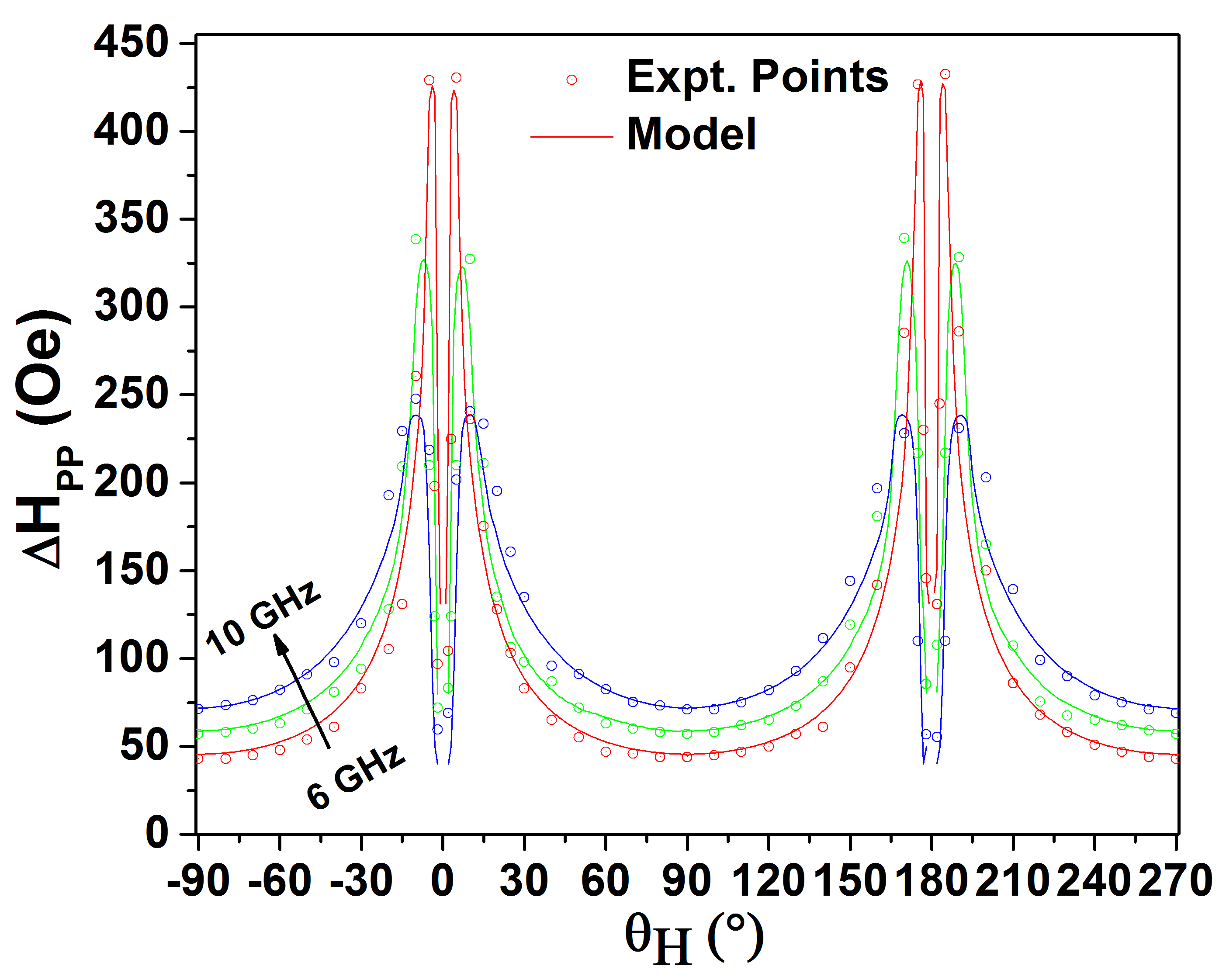}\
\caption{Out-of-plane angular dependence of total linewidth, $\Delta H_{pp}$ at different frequencies. $(\circ)$ shows the experimental points and the line $(-)$ shows the modeled data.}\label{F5}
\end{center}
\end{figure}

\begin{equation}
\begin{array}{c}
\Delta {H_{PP}} = \Delta H(\alpha ) + \Delta H\left( {\Delta 4\pi {M_{eff}}} \right) + \Delta H\left( {\Delta {\theta _H}} \right)\\
 = \frac{2}{{\sqrt 3 }}\frac{1}{{\left| {\frac{{\partial \omega }}{{\partial {H_r}}}} \right|}}\frac{{\alpha \gamma }}{{{M_S}}}\left( {\frac{{{\partial ^2}E}}{{\partial \theta _M^2}} + \frac{1}{{{{\sin }^2}{\theta _M}}}\frac{{{\partial ^2}E}}{{\partial \varphi _M^2}}} \right) + \\
\frac{1}{{\sqrt 3 }}\left( {\left| {\frac{{\partial H}}{{\partial 4\pi {M_{eff}}}}} \right|\Delta 4\pi {M_{eff}}} \right) + \frac{1}{{\sqrt 3 }}\left( {\left| {\frac{{\partial H}}{{\partial {\theta _H}}}} \right|\Delta {\theta _H}} \right)
\end{array}\
\end{equation}
where $4\pi M_{eff}=4\pi M_{S}-2K_{\bot}/M_{S}$, is the effective magnetization. $\Delta H(\alpha)$ arises from the intrinsic Gilbert type damping and has large contribution towards linewidth broadening. The parameter $\alpha$ signifies how fast the precessional energy is dissipated into the lattice. The terms $\Delta H(4\pi\Delta M_{eff})$ and $\Delta H(\Delta\theta_{H})$ represent the linewidth broadening due to the spatial dispersion of the magnitude and direction of $M_{eff}$, respectively. The $\theta_{H}$ dependence of $\Delta H_{PP}$ was modeled by using Eq. 6 and the interpolated values of $\theta_{M}$ from Fig. 3 at different frequencies. The numerically calculated values of $\Delta H_{PP}$  are shown as solid lines in Fig. 5. The individual contributions towards the total linewidth ($\Delta H_{pp}$) is also shown in Fig. 6.
\begin{figure}[t]
\begin{center}
\includegraphics[trim = 0mm 0mm 0mm 0mm, width=80mm]{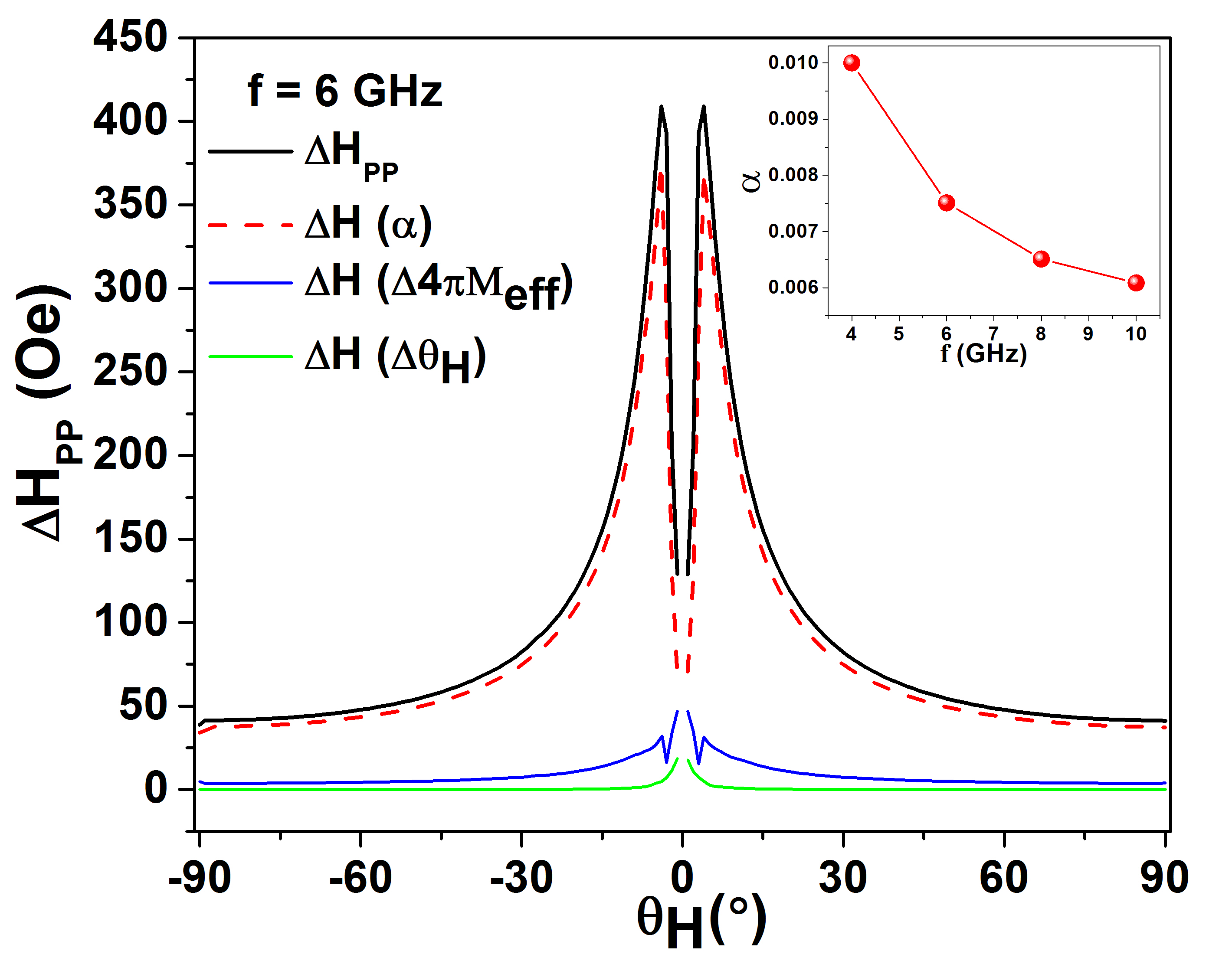}\
\caption{$\theta_{H}$ dependence of $\Delta H_{pp}$, $\Delta H(\alpha)$, $\Delta H(\Delta4\pi M_{eff})$ and $\Delta H(\Delta\theta_{H})$ modeled data at 6 GHz frequency.}\label{F6}
\end{center}
\end{figure}
The curves  are shown for a single frequency for better clarity. The linewidth broadening is observed mainly due to the intrinsic Gilbert damping. The extrinsic contribution is found to be negligible when $\theta_{H}$ is away from $0^{\circ}$ and $180^{\circ}$ but it is large near the perpendicular configuration. The Gilbert damping parameter at different frequencies for the FeTaC thin film is plotted in the inset of Fig. 6. It shows that the $\alpha$ decrease monotonically with the increase in frequency. The low value of damping parameter observed in the present thin film can be more relevant towards the STT technology or MTJ applications. Such an increase of $\alpha$ by decreasing precessional frequency has been attributed to the inhomogeneous linewidth broadening due to the dispersion of anisotropic field.\cite{Celinski1991,Silva1999} The dispersion in magnitude and direction of effective magnetization are found to be $\Delta 4\pi M_{eff}$=0.1 KOe, $\Delta \theta_{H} \approx 1\times10^{-4}$ degree. The value of Gilbert damping constant for the present FeTaC thin film is found to be comparable to those reported in Fe-based magnetic thin films, such as FePd ternary alloy\cite{He2013}, permalloy\cite{Counil2004}, NiFe/CoFeB/CoFe multilayered sturucture\cite{Lu2014} and (FeCo)$_{1-x}$Gd$_{x}$\cite{Guo2014}. However, the Mn- and Co- based  thin films\cite{Mizukami2013,Song2013} have larger damping parameter as compared to the present film which could be understood on the basis of spin-orbit coupling.

In conclusion, PMA and Gilbert damping which are very important and crucial parameters for STT, STT-MRAM and TMR applications, have been analyzed in FeTaC soft ferromagnetic thin film with a striped domain structure by using MS-FMR technique in broad band frequency range. The precise estimation of Land\'{e} $g$-factor, PMA constant and $4\pi M_{eff}$ were carried out by using total energy density function for magnetic thin film. Spin dynamics relaxation which is quantified by Gilbert damping parameter has been analyzed at different frequencies and is found to be 0.006 which falls in the most reliable order from application point of view. The values of $\alpha$ are found to be comparable to those reported Fe-based single layer and multilayered magnetic thin films.\\

\textbf{Acknowledgement}\\

The authors would like to thank Dr. Najmul Haque for helping to write energy minimization calculation in mathematica and Mr. Nazir Khan for his help during writing data acquisition software.

\bibliography{FMR}

\end{document}